\documentclass[floatfix,aps,prl,twocolumn,groupedaddress,showpacs,english]{revtex4-2}\usepackage{graphicx} 
\usepackage{amsmath}
\usepackage{soul}
\usepackage{subcaption}
\usepackage{graphicx}
\usepackage{xcolor} 

\makeatother
\renewcommand{\hl}[1]{#1}   

\begin{document}
\title{\hl{Driven Atoms and Molecules as Coherent Attosecond Waveform Processors}}
\author{Asaf Farhi}
\affiliation{
      School of Physics and Astronomy, Faculty of Exact Sciences, Tel Aviv University, Tel Aviv 69978, Israel}\vspace*{2cm}

\begin{abstract}
Advancing temporal resolution in computation, signal modulation, and quantum measurement is \hl{fundamentally constrained in optical resonator platforms by a trade-off between operation time and temporal resolution}. We show that driven atomic and molecular systems perform temporal integration of resonant pulses \hl{via passive coherent absorption or stimulated emission}, enabling attosecond resolution together with long operation time. We derive an analytic description of this mechanism through Bloch equations and time-dependent Schrödinger equation, finding quantitative agreement. We further identify feasible atomic transitions and excitation schemes accessible with current technology, and describe implementations for temporal differentiation and waveform generation at attosecond resolution.  These results establish driven quantum transitions as a \hl{framework} for attosecond-scale waveform processing beyond conventional resonator \hl{approaches}, with potential applications in computation, \hl{optical switching}, system control, and data transfer. 

\end{abstract}
\maketitle

\paragraph*{Introduction}The temporal resolution of signal processing and modulation is of fundamental importance for fast computation, high-rate data transmission, optical switching, and the study of ultrafast phenomena \cite{mohammadi2019inverse}. Standard electronic devices achieve modulation frequencies on the order of a gigahertz, but advancing these domains requires substantially faster temporal control. In particular, enhancing the temporal resolution can enable performing highly complex computations on ultrashort timescales. Moreover, fast signal modulation can allow  secure communication at very high speeds, waveform synthesis with ultrafine temporal control, and rapid optical switching, a cornerstone of photonic signal processing \cite{venkataraman2011few,yavuz2006all,dawes2005all}. At the same time, precise shaping and control of ultrashort pulses are required to probe ultrafast dynamics—from electron motion in atoms and molecules to processes in condensed matter—recently demonstrated through high-harmonic generation \cite{uzan2024observation}. Thus, pushing the limits of temporal resolution can unlock new capabilities across computation, communication, photonics, and fundamental physics, driving both science and technology forward.



A key operation in both computation and signal modulation is real-time integration, enabling applications ranging from solving differential equations, to molecular dynamics, fluid dynamics, astrophysics, and quantum chemistry \cite{mohammadi2019inverse}.
 Another fundamental operation is differentiation \cite{asghari2008proposal,slavik2006ultrafast,sol2022meta}, which plays a central role in machine learning, optimization, computational physics, and chemistry. Both of these operations are at the heart of control theory, with controllers that utilize them to bring physical systems to a desired state \cite{sutskever2013importance,johnson2005pid}. Despite recent advances, these operations often restrict efficiency and present a bottleneck in diverse fields.
Recently, significant interest has emerged in using optical resonators for computation and signal modulation, where the temporal resolution is determined by the round-trip time and the operation time by the resonator lifetime. Active cavities such as a laser at threshold, which exhibit a real-frequency pole, have been utilized to perform integration of the field envelope up to the gain saturation time \cite{slavik2008photonic}.
Passive cavities with loss but no gain have been shown to perform an operation that is similar, but distinct, from integration of real-frequency pulses with no energy consumption \cite{ferrera2010chip}. Similarly, cavities tuned to coherent perfect absorption can differentiate signals at optical frequencies \cite{chong2010coherent,wan2011time,farhi2022excitation,sol2022meta}. More recently, a theoretical work demonstrated that a short-cavity laser can achieve a modulation frequency one order of magnitude below the carrier frequency, with picosecond operation times \cite{farhi2024generating}. Nevertheless, such resonators face intrinsic trade-offs: increasing the resonator size extends the operation time but inherently limits the achievable temporal resolution, as is shown in Fig. 1 (a). Furthermore, synthesizing short cavities is rather challenging and requires distributed Bragg reflector designs to increase the Q factor, fundamentally limiting the temporal resolution \cite{farhi2024generating,slavik2008photonic,ferrera2010chip,wan2011time,farhi2022excitation}.  
\begin{figure*}[!t]
    \centering
    \includegraphics[width=18cm]{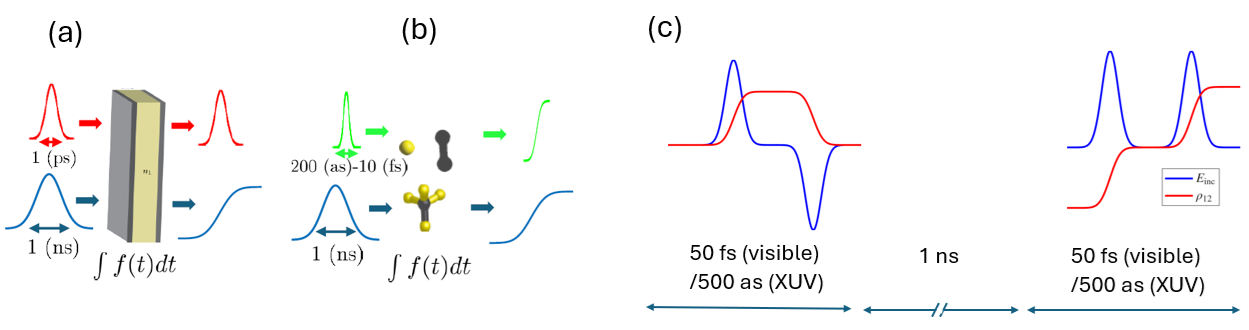}
    \caption{Optical resonators and atoms in waveform processing. (a) A typical slab resonator with a thickness of 1000$\lambda$ exhibits picosecond resolution and operates for 1 ns. This enables integrating 1 ns pulses but results in scattering of the original 1 ps incoming pulse. Temporal integration of short pulses is limited by the round-trip time, which sets a boundary on the response speed. Shortening the slab thickness improves the temporal resolution but reduces the operation time. (b) Atoms have dramatically improved performance with attosecond resolution and operation time of 1 ns and even longer. This temporal resolution enables them to integrate Gaussian pulses of 200 (as)-10 (fs), corresponding to UV and visible frequencies, which requires 10 (as) and 500 (as) resolution, respectively, since $\int f\left(t\right)\approx\sum_{i}f_{i}\Delta t$, and also 1 ns pulses. (c)  The combination of exceptionally high temporal resolution and long operation times in atoms allows for attosecond processing of substantial amounts of data.}
    \label{fig:enter-label4}
\end{figure*}
\vspace{-2pt}

While these works focused on real-frequency excitations, recent research has explored resonant complex-frequency excitations in passive resonators across diverse physical platforms, opening promising frontiers in wave physics and device engineering. Such excitations have been analyzed classically in the frequency domain and shown to produce exotic phenomena, including overcoming loss in superlenses, enhanced propagation of phonon polaritons, and surpassing conventional scattering limits \cite{kim2022beyond,kim2023loss,guan2024compensating,guan2023overcoming}. Very recently, the author and colleagues studied the time-domain response to resonant complex-frequency excitation of subwavelength structures, which have improved temporal resolution due to their short roundtrip. They derived closed-form expressions for the temporal response of such resonators, generalized these results to complex-frequency exceptional points, and experimentally demonstrated both the theoretical predictions and superior power efficiency \cite {farhi2025time}. However, so far the highest Q factor obtained for subwavelength structures is 450 \cite{herzig2024high}, limiting their operation time to approximately 40 optical cycles. 

Here we show that coherently driven atoms and molecules enable a previously inaccessible combination of attosecond temporal resolution and long operation times, arising from coherent light–matter dynamics in a deeply subwavelength quantum system with long-lived coherence. We find that their \hl{absorption or stimulated emission} response to resonant arbitrary pulse excitations is the integral of the incoming-field envelope (Fig. 1 (b)). Importantly, this allows processing of long pulses with high modulation frequencies, with direct implications in computation, modulating signals, system control, and optical switching (Fig. 1 (c)). Interestingly, as we consider in our analysis two-level systems, it applies to various effectively-two-level atomic and molecular transitions including electronic, vibrational, rotational, and spin, while maintaining this attosecond resolution for localized transitions. We demonstrate our results theoretically for atoms using Bloch equations and time-dependent Schrödinger equation (TDSE), showing excellent agreement with analytical integration, and compare them with the standard slab resonator. Finally, we propose techniques for pulse generation, differentiation, and attosecond optical switching, together with experimentally feasible implementations, opening the way to a broad class of attosecond operations.

\paragraph*{Results}
Our analysis applies to arbitrary resonant pulse excitation with a carrier frequency equal to the transition frequency. We begin by considering a complex-frequency drive of an atom initially in a ground or excited state,
$E(t) \propto \theta(t) \, e^{i \omega t - \Gamma t},$
where $\Gamma$ is the atomic decay (coherence) rate, $\omega$ is the transition frequency, and $\theta(t)$ is the Heaviside function. We assume the short-time limit, $t \ll 1/\Gamma$, so that the drive reduces to $E(t) \approx \theta(t) \, e^{i \omega t}.$
We also assume that the pulse temporal width satisfies $t\ll 1/\Omega,$ where $\Omega$ is the Rabi frequency, a regime common in transient absorption experiments \cite{scully1997quantum,goulielmakis2010real}.  For simplicity, we first focus on two-level systems; however, as shown in the Supplementary Material (SM), atoms with split energy levels—including those arising from Rabi splitting—and well-separated excited states can often be treated effectively as two-level systems in practical cases.
We employ the analytic two-level Bloch equations for arbitrary field amplitude in Refs. \cite{alekseev1992analytic,Alekseev1990}, formulated within the dipole approximation. Such Bloch equations have been successfully employed for attosecond pulses in atoms and semiconductors \cite{arkhipov2024generation,chen2025attosecond}. 
We calculate the coherent absorption or stimulated emission via the expression for the polarization derived from the off-diagonal density-matrix element $\rho_{12}$.
\begin{widetext}
\begin{gather}
\mathrm{Im}(\rho_{21})=\frac{n_{0}\Omega e^{-\Gamma t}\left[\left(\text{Ci}\left(\frac{\text{\ensuremath{\Omega}}}{\Gamma}\right)-\text{Ci}\left(\frac{e^{-\Gamma t}\Omega}{\Gamma}\right)\right)\cos\left(\frac{\Omega e^{-\Gamma t}}{\Gamma}\right)+\left(\text{Si}\left(\frac{\Omega}{\Gamma}\right)-\text{Si}\left(\frac{e^{-\Gamma t}\text{\ensuremath{\Omega}}}{\Gamma}\right)\right)\sin\left(\frac{\Omega e^{-\Gamma t}}{\Gamma}\right)\right]}{\Gamma}\approx\nonumber\\n_{0}\Omega t\left[1-\Gamma t-\frac{1}{6}t^{2}\left(\Omega^{2}-3\Gamma^{2}\right)\right]+O\left(t^{4}\right),\,\,
P=\rho_{21}e\left\langle a|x|b\right\rangle e^{i\omega t}+c.c
=2e\mathrm{Im}(\rho_{21})\left\langle a|x|b\right\rangle \sin\left(\omega t\right),
\end{gather}
\end{widetext}
\begin{figure*}
    \centering
    \includegraphics[width=17cm]{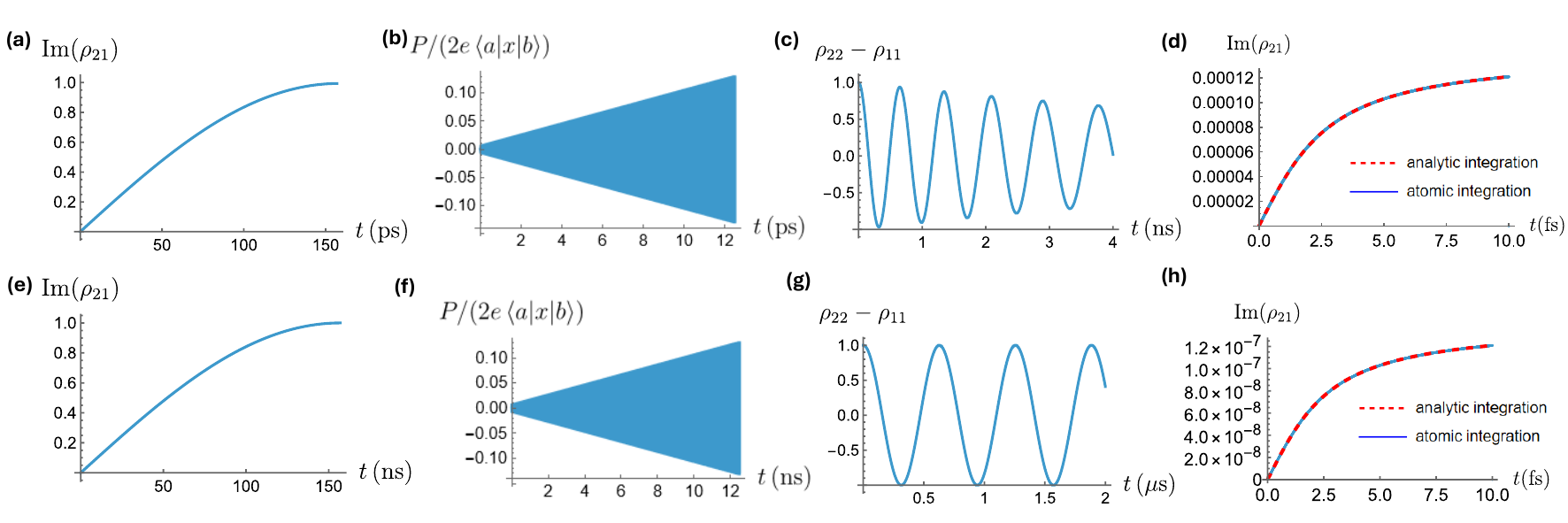}
    \caption{The response of two atoms to external excitations of $E\propto \cos(\omega t)\exp(-\Gamma t)$ and $E\propto f(t) \cos(\omega t)$. The response of an atom with $\omega_\mathrm{ge}=3.76\cdot 10^{15}\,\mathrm{(1/s)},\,\,\Gamma=10^8\,\mathrm{(1/s)},\,\,\mathrm{and}\,\,\Omega_R=10^{10}\,\mathrm{(1/s)}$  to external excitations: (a) Im$(\rho_{12}(t))$ (b) $P(t)$ and (c) $\rho_{22}(t)-\rho_{11}(t)$ for the excitation $E\propto \cos(\omega t)\exp(-\Gamma t)$. The response of $\propto t\exp(i\omega t)$ in (b) is a signature of a pole in the transfer function of the atom.  (d) Im$(\rho_{12}(t))$ in response to the excitation $E\propto (1/(1+t^2))\cos(\omega t)\exp(-\Gamma t)$ or $E\propto (1/(1+t^2))\cos(\omega t),$ which is in very good agreement with its integral $\arctan(t).$ Due to the high Q factor of the atom, the effect of $\exp(-\Gamma t)$ in the input at short times is negligible. We also calculate these quantities for an atom with $\Omega=10^7\,(\mathrm{1/s})$ and $\Gamma=1\,(\mathrm{1/s})$ in (e)-(h). As can be seen in (f) and (h) the pole regime time lasts more than $10\,\mathrm{ns}$ and the ultrahigh temporal resolution is maintained throughout this time, which is on the order of $4\cdot 10^7$ optical cycles. Here, too, there is excellent agreement between the analytic integration and the atomic integration.}
    \label{fig:enter-label2}
\end{figure*}
where $\text{Ci}\left(t\right)=-\intop_{t}^{\infty}\frac{\cos t'}{t'}dt',\,\,\,\text{Si}\left(t\right)=-\intop_{t}^{\infty}\frac{\sin t'}{t'}dt',$ $n_0=\rho_{22}(t=0)-\rho_{11}(t=0),$ and we considered the first orders in the expansion. Interestingly, this change between the drive and atom polarization of $1\rightarrow t$ implies a pole in the atom transfer function and that the atom polarization is the integral of the incident field envelope with no temporal resolution limit under the dipole approximation; we verify this statement using the full expression in Eq. (1) and in the SM with additional directions including TDSE. When an atom is initially in ground or excited state, this induced polarization, which has a $\pi/2$ phase with respect to the excitation, produces absorption or stimulated emission that is directly observable (see analysis for absorption in Ref. \cite{goulielmakis2010real}). Note that circular-polarization excitation eliminates counter-rotating waves, making this result very accurate \cite{scully1997quantum} for short pulses, with superachromatic-waveplate implementation. Note also that this calculation in the dipole approximation neglects the field propagation in $z.$ This means that the allowed temporal resolution is on order of 10 attoseconds for angstrom-size wavefunction extent as derived in detail in the SM. Molecular transitions are also often localized in an angstrom scale, as in the cases of vibrational single-bond transitions, spin transitions, and rotational transitions of small molecules \cite{farhi2024giant}. A standard way of initializing a two-level system in excited state is via a $\pi$ pulse. \hl{In absorption mode, typically there is no need to initialize the system and invest energy as atoms and molecules are often in a ground state. Since initializing an atom in an excited state is rather challenging at ultraviolet frequencies, absorption-mode operation is more straightforward and allows exploitation of the 10-attosecond temporal resolution enabled by these frequencies.}.  
 To ensure that the density matrix dynamics will be dominated by the drive, we also consider field intensity where $\Omega\gg \Gamma$.

The atom polarization can be measured via attosecond or femtosecond pump-probe spectroscopy  \cite{ goulielmakis2010real,kneller2025attosecond} or by detecting the absorbed or stimulated emission field. Analysis of the radiated field of point dipoles with a general time dependence shows that $E_\mathrm{rad}(\mathbf{x},t)\propto \frac{e^{ikr}}{r} d^2p(t)/dt^2$ since $\boldsymbol{A}\propto \frac{e^{ikr}}{r}\frac{dp(t)}{dt}$ and $\boldsymbol{E}_\mathrm{rad}\propto d \boldsymbol{A}\left(t\right)/dt,$ where $\mathbf{A}$ is the magnetic vector potential \cite{jackson2021classical}. As $p(t)\propto e^{i\omega t}\int E_\mathrm{env}dt',$ we obtain $E_\mathrm{rad}(\mathbf{x},t)=\frac{e^{ikr}}{r}\left(\frac{dE_\mathrm{env}}{dt}e^{i\omega t}+2i\omega E_\mathrm{env}e^{i\omega t}-\omega^2e^{i\omega t}\int E_\mathrm{env}dt'\right).$ To obtain pure integration, one has to subtract both the proportional and derivative signals as we explain in the SM.\hl{ In the  forward-propagating pulse regime, which is typically employed in bulk media, the governing equation reduces to $\frac{\partial E}{\partial z}\approx -\frac{2\pi}{c}\frac {\partial P}{\partial \tau},\,\,\tau = t - \frac{z}{c}$ , where we used the plane-wave approximation } \cite{geissler1999light}. \hl{Since $E(z,t) \propto
e^{i\omega \tau}
\left[
E_{\mathrm{env}}(\tau)
+ i\omega \int_{-\infty}^{\tau}
E_{\mathrm{env}}(\tau')\, d\tau'
\right],$ it enables to obtain pure integration by subtracting only $E_{\mathrm{inc}}= E_\mathrm{env}e^{i\omega t}$ from the resulting signal, similarly to standard interference experiments. Note that we first solved the Bloch equations to calculate the induced atomic polarization, and subsequently determined the generated field. This sequential approach is justified in the weak-signal regime, where the ratio of the generated field to the incoming field is negligibly small.} We also propose that using an attosecond delay, which can be realized by a translation stage with attosecond precision \cite{chen2025attosecond}, and superachromatic half-wavelength plate to generate $[f(t+\Delta t)-f(t)]/\Delta t$, one could obtain the derivative of the field $E_\mathrm{output}(x,t)=\frac{dE_\mathrm{env}}{dt}e^{i\omega t}+i\omega E_\mathrm{env}e^{i\omega t}$ with attosecond resolution.
These integration and differentiation operations can be used for modulating and demodulating signals e.g., in data transmission or for Proportional–Integral–Derivative (PID) controllers, which are widely used in engineering for stabilizing and controlling dynamic systems. 

To demonstrate this behavior, we first consider an atom with the typical values of $\Gamma=10^8\, (\mathrm{1/s}),\,\,\Omega=10^{10}\, (\mathrm{1/s}),\,\,\omega=3.76\cdot 10^{15}\, (\mathrm{1/s})$ and resonantly excite it with a pulse of $E_\mathrm{inc}=\theta(t)e^{-\Gamma t}\cos(\omega t)\approx \theta(t) \cos(\omega t)$. In Fig. 2 (a) and (b) we present Im$[\rho_{12}(t)]$ for the pulse duration and $P(t)/(2e\left\langle a|x|b\right\rangle)$ for $t\ll 1/\Omega$, respectively, from the full expressions in Eq. (1). Interestingly, the polarization envelope in (b) is proportional to $t,$ which implies that there is a pole behavior in the polarization. In addition, this behavior lasts for at least 400000 optical cycles, with a ratio of pulse duration to temporal resolution of $10^{7}.$  In Fig. 2 (c) we present the inversion $\rho_{22}-\rho_{11}$, which oscillates at long times as expected but its amplitude is decreasing as opposed to cw drive.
To verify the integration operation of the atom, we numerically calculated the polarization from the full expression in Eq. (1) for an incident field of the form $E\propto \theta(t) \frac{4}{1+t^{2}}\cos(\omega t)\exp(-\Gamma t)\approx \theta(t)\frac{4}{1+t^{2}}\cos(\omega t),$ which is similar to the Gaussian pulse in Fig. 1 (b), and we expect the polarization envelope to be equal to $\intop_{-\infty}^{t}4/(1+t'^{2})dt'=4\arctan\left(t\right)$. Indeed, the result of Im$(\rho_{21})$ in Fig. 2 (d) agrees very well with this expression and underscores the ultrahigh temporal resolution of the atom operation as this 10 fs pulse can be decomposed to tens of time steps, which requires sub fs resolution. Note that this 10 fs was chosen due to the Gaussian pulses at hand and one could also consider shorter pulses on the order of 200 (as) at higher UV frequencies \cite{chen2025attosecond,kneller2025attosecond}. More generally, such behavior is expected for any pulse shape including a pulse train as long as $t\ll 1/\Omega,$ as depicted in Fig. 1(c). It is worth noting that due to the very high Q factors of atoms and molecules, the drive is not required to have the $\exp(-\Gamma t)$ part of the function, considerably simplifying the operation with any input of the form $f(t)\exp(i\omega t),$ where $f(t)$ is integrated in $P(t).$

 We then consider an atom or molecule with $\Omega\sim 10^7\, (\mathrm{1/s})$ and $1/\Gamma\sim \mathrm{s}$ \cite{katz2018light}, where the pole regime lasts for $4\cdot 10^{7}$ optical cycles and the pulse-duration-to-temporal-resolution ratio is $\sim 10^{9}$. On timescales shorter than both $1/\Omega$ and $1/\Gamma$, the system behaves effectively linearly. In Fig. 2 (e)-(h) we plot Im$(\rho_{21}),\,\,P,\,\,\rho_{22}-\rho_{11},$ and the integration result for the above pulse, respectively. Strikingly, the pole regime lasts for more than 10 ns and the same ultrahigh temporal resolution is maintained, albeit with a lower polarization signal. To process sub-fs pulses one can excite UV or extreme UV transitions so that a pulse will include several cycles. Such transitions occur for example in  $\mathrm{He}, \mathrm{He}^+$ having $\Gamma\sim 1 \,\mathrm{(ns)}$ with a concrete analysis in the SM \cite{kneller2025attosecond}.

To analyze the effect of the classical-resonator size on the temporal resolution, we analytically calculate the response of a passive (without gain) large-than-wavelength one-sided slab with a perfect mirror on one side. The total reflection coefficient \cite{farhi2024generating} and the complex-frequency drive in the time domain read
\[
r=-\frac{r_{1}+e^{2i\frac{\omega}{c}nl_{1}}}{1+r_{1}e^{2i\frac{\omega}{c}nl_{1}}},\,\,y_{1}=e^{-\Gamma t+i\omega t}\theta\left(t\right),
\]
where $r_1$ is the reflection coefficient of the slab, $l_1$ is the slab length, $n$ is its refractive index, and $\omega,\Gamma$ of the resonance were calculated by imposing vanishing of the denominator. Since the denominator is a sum of a geometric series, we expand it accordingly in time and convolute the result with the input to obtain
\begin{gather}
d\left(t\right)=\delta(t)-r_{1}\delta\left(t-2\frac{nl_{1}}{c}\right)+r_{1}^{2}\delta\left(t-4\frac{nl_{1}}{c}\right),\,\,y_{1}*d\nonumber\\
=e^{-\Gamma t-i\omega t}\left[\theta(t)-r_{1}e^{\Gamma\left(2\frac{nl_{1}}{c}\right)+i\omega\left(2\frac{nl_{1}}{c}\right)}\theta\left(t-2\frac{nl_{1}}{c}\right)+...\right]
\nonumber\\=e^{-\Gamma t-i\omega t}\left[\theta(t)+\theta\left(t-2\frac{nl_{1}}{c}\right)+...\right],
\end{gather}
where in the last transition we substituted the resonance condition. 
Incorporating the second term in the numerator, we get a similar expression
delayed by $2\frac{nl_{1}}{c}:$
\[
E_\mathrm{scat}=\left[r_{1}+\delta\left(t-2\frac{nl_{1}}{c}\right)\right]*e^{-\Gamma t+i\omega t}\left[\theta(t)+\theta\left(t-2\frac{nl_{1}}{c}\right)+...\right].
\]
One can readily see that the discretization of the response arises from the cavity roundtrip, which highlights the fact that using small resonators improves the resolution.
 To compare the slab performance to the first atom, we consider a slab with the same operation time in optical cycles, having the following parameters:  $l_1=1800\lambda,\,\,r_1=0.99,\,\,n=1.4,\,\mathrm{Q}\,\,\mathrm{factor}=3.68\cdot 10^6,$ and $\lambda=550\mathrm{nm}.$ 

 In Fig. 3 (a) we plot the scattered field for an incoming field with a complex-frequency excitation of the form $e^{-\Gamma t-i\omega t}$ for $t\ll 1/\Gamma$. Evidently, the response is composed of steps with a time difference of the roundtrip of the slab of $\Delta t=10\mathrm{ps,}$ with a temporal resolution 4 and 5 orders of magnitude lower than the atom for visible and UV frequencies, respectively. Crucially, when we input the same pulse as in Fig. 2 (d) of $e^{i\omega t}\frac{1}{1+t^2}$ for 10fs, we get that the scattered field is the incoming field multiplied by $r_1$ i.e., $r_1e^{i\omega t}\frac{1}{1+t^2}$ (did not perform any integration) instead of its integral due to the convolution with only $\delta(t)$ during this pulse as shown in Fig. 1 (a), where we neglected the exponential since $\exp(-\Gamma t)\approx 1$. This agrees with the fact that only the first reflection from the slab interface comes into play during this time and highlights the fundamentally superior performance of atomic and molecular integration. In Fig. 3 (b) we plot the response of the slab for longer times. Clearly, at longer times $E_\mathrm{scat}$ behaves similarly to the response of the classical subwavelength passive resonators considered in Ref. \cite{farhi2025time} and an atom with $\Gamma\gg \Omega$. Here the linear behavior lasts for approximately the same number of optical cycles as in the case of the first atom. An analysis of a short slab is given in the SM.

\begin{figure}[htbp]
  \centering
  \begin{subfigure}[b]{0.45\textwidth}
\caption{}
    \includegraphics[width=\linewidth]{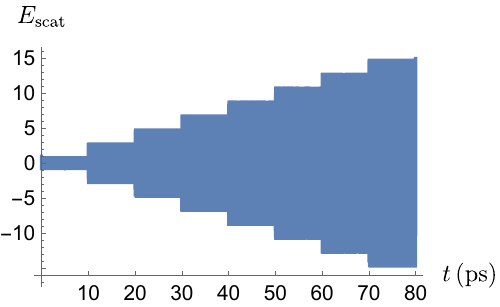}
    \label{fig:sub1}
   
  \end{subfigure}
  \hfill
  \begin{subfigure}[b]{0.45\textwidth}
 \caption{}  
    \includegraphics[width=\linewidth]{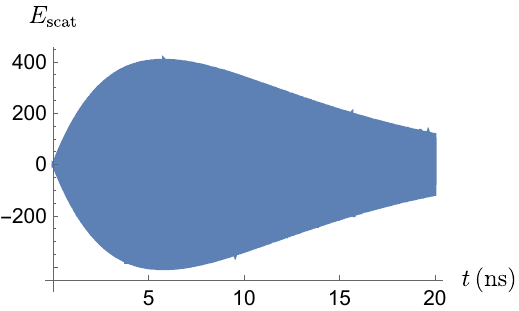}
    \label{fig:sub2}

  \end{subfigure}
  \caption{The response of a lossless slab to $E_\mathrm{inc}=e^{-\Gamma t+i\omega t}$. (a) The initial response of a slab with $l=1800\lambda, r_1=0.99,\,\, n_1=1.4,\lambda=550\mathrm{nm},\,\,\mathrm{and}\,\, Q=3.68\cdot 10^6$ to the complex-resonant-frequency incoming field $E_\mathrm{inc}=\exp(-i\omega t -\Gamma t)\theta(t).$ Unlike the previous case, here the long roundtrip results in  discretization in the response, which reduces the temporal resolution by more than $10^6$. (b) The scattered field for the complex-frequency excitation for  $t<4/\Gamma,$  of the form $E_\mathrm{scat}\propto t\exp(-i\omega t -\Gamma t)\theta(t).$}
  \label{fig:main}
\end{figure}


Finally, we propose two approaches to generate pulses with attosecond modulation frequencies. Such pulses are required for performing arithmetic operations at ultrahigh resolution as well as for a variety of purposes, including high-rate data transmission, secure communication, and probing ultra-fast phenomena. We start with the transform-limited femtosecond pulses at hand, which are Gaussian.  Differentiation of the envelope can be performed with attosecond resolution by employing translation stages with attosecond precision  \cite{chen2025attosecond}, followed by interference with the wave itself as proposed above. We can proceed in this manner to generate the high-order Hermite-Gaussian modes \cite{asghari2008proposal} given by  $\psi_m(x)=-1^m\frac{d^m}{dx^m}e^{-x^2}$ via $\frac{d}{dt}e^{-(\xi/\tau)^{2}}e^{i\omega \xi}=-2\frac{\xi}{\tau^2} e^{-(\xi/\tau)^{2}}e^{i\omega \xi}+i\omega e^{-(\xi/\tau)^{2}}e^{i\omega \xi}\overset{\mathrm{interference}}{\rightarrow}-2\frac{\xi}{\tau^2} e^{-(\xi/\tau)^{2}}e^{i\omega \xi},$
$\frac{d}{dt}2\frac{\xi}{\tau^2} e^{-(\xi/\tau)^{2}}e^{i\omega \xi}=f'g+fg'\overset{\mathrm{interference}}{\rightarrow}f'g ,$ where $\xi=t-x/c,$ $f=2\frac{\xi}{\tau^2} e^{-(\xi/\tau) ^2},
\,\,g=e^{i\omega\xi},$ and the second interference was generated using the output of the first. Thus, we can perform consecutive differentiations to generate this set of functions that span space. To construct any desired function $h(x)$ at ultrahigh modulation frequencies, we can utilize a superposition of these orthogonal Hermite-Gaussian modes, where the coefficients are given by $c_m=\int \psi_m(x') h(x')dx'.$  A second approach to generate desired pulses at ultrahigh modulation frequencies would be to integrate a window or step function multiplied by $e^{i\omega t}$ consecutively with the corresponding weights so that one could construct a pulse using a Taylor series expansion. While this has been discussed in Ref. \cite{farhi2024generating} in the context of optical cavities, here our integrator is an atom or molecule, enabling significantly higher modulation frequencies. This 
approach entails the challenge of generating the initial step or window function at such high modulation frequencies. Fortunately, an atom naturally transforms femtosecond Gaussian pulses to such functions, offering a straightforward approach for femtosecond optical switching as shown in Fig. 1 (b)  \cite{venkataraman2011few,yavuz2006all,dawes2005all}.

\paragraph*{Discussion}
We showed that two-level systems can process waveforms \hl{via coherent absorption or stimulated emission}, achieving attosecond resolution for localized transitions while maintaining long operation times. We demonstrated this for electronic transitions in atoms using Bloch equations and the time-dependent Schrödinger equation (TDSE), showing excellent agreement with analytical calculations. Compared with standard larger-than-wavelength resonators, these systems offer significantly enhanced temporal resolution and figure of merit defined as the ratio of operation time to temporal resolution. The mechanism is broadly applicable to molecules and to various types of transitions, including electronic, vibrational, rotational, and spin~\cite{farhi2024giant}. We proposed techniques to differentiate and synthesize arbitrary pulses at attosecond resolution, and to implement femtosecond optical switching. Unlike standard pulse-modulation schemes, which are tailored to specific pulse shapes and limited in temporal resolution, our approach works for arbitrary waveforms and has significantly improved termporal resolution. These operations also enable a wide range of \hl{functionalities} such as multiplication, division, solving differential equations~\cite{mohammadi2019inverse}, and modulation/demodulation schemes. Experimental realization of these phenomena is feasible using standard atomic localization and control methods, including traps, optical lattices, gas cells, two-dimensional atom arrays, and potentially surface adsorption  \cite{demille2002quantum,finkelstein2024universal,bekenstein2020quantum,sugimoto2007chemical,levin2025coherent,kneller2025attosecond}.  Observing this phenomenon can be achieved through pump-probe transient absorption / interferometry or pulse characterization \cite{goulielmakis2010real,kneller2025attosecond,pichard2024rearrangement,gyger2024continuous}. Closely spaced excited levels can be effectively treated as a single two-level system, preserving the temporal integration behavior. Using this principle, we identify several atomic species — Rb, Sr, He, and $\mathrm{He^+}$ — with transition wavelengths and level structures suitable for processing pulses with femtosecond or attosecond resolution (SM). \hl{Our results position atomic systems as a transformative platform for attosecond-resolution, time-domain quantum processing, suggesting a pathway to computations and signal modulation that are up to $10^5$ and $10^8$ times faster than optical-resonator-based techniques and conventional computers, respectively.} Future research could focus on pumped gain media in micro and nano devices, potentially enabling to keep the initial-state population constant for larger $\Omega$ values, thereby increasing the emitted-field signal. \hl{Alternatively, in absorption mode micro or nano devices doped with rare earth ions in ground state combine high efficiency and passive behavior.  }

 \paragraph*{Acknowledgement}
 H. Suchowski, N. Dudovich, R. Finkelstein, E. Shahmoon, G. Marcus, D. Oron, Y. Israel, S. Korenblit,  and Y. Piasetzky are acknowledged for their insightful comments and valuable suggestions.

\bibliographystyle{unsrtnat}
\bibliography{bib2}
\clearpage

\begin{widetext}
\section*{Supplementary Material}

\setcounter{figure}{0}
\subsection{Analysis of a system with two closely-spaced excited states}
It is often the case that  excited and ground states of an atom or molecule are split in frequency to several closely-spaced hyperfine energy
levels \cite{ye1996hyperfine}. Now we show that for short pulses  and small splitting, such atoms and molecules can be treated as two-level systems.
We write the Bloch equations for a three-level system with a ground state $a$ and excited states $b,d$:
\begin{gather}
\dot{C}_{a}=-i\omega_{a}C_{a}+i\Omega_{ab}e^{-i\phi_{ab}}\cos\left(\nu t\right)C_{b}+i\Omega_{ad}e^{-i\phi_{ad}}\cos\left(\nu t\right)C_{d},
\nonumber\\
\dot{C}_{b}=-i\omega_{b}C_{b}+i\Omega_{ab}e^{i\phi_{ab}}\cos\left(\nu t\right)C_{a}+i\Omega_{bd}e^{-i\phi_{bd}}\cos\left(\nu t\right)C_{d},
\nonumber\\
\dot{C}_{d}=-i\omega_{d}C_{d}+i\Omega_{ad}e^{i\phi_{ad}}\cos\left(\nu t\right)C_{a}+i\Omega_{bd}e^{i\phi_{bd}}\cos\left(\nu t\right)C_{b}.\nonumber
\end{gather}
We substitute
$$
C_{a}=c_{a}e^{-i\omega_{a}t},\,\,C_{b}=c_{b}e^{-i\omega_{b}t},\,\,C_{d}=c_{d}e^{-i\omega_{d}t},
$$
and obtain:
\begin{gather}
\dot{c}_{a}e^{-i\omega_{a}t}=i\Omega_{ab}e^{-i\phi_{ab}}\cos\left(\nu t\right)c_{b}e^{-i\omega_{b}t}+i\Omega_{ad}e^{-i\phi_{ad}}\cos\left(\nu t\right)c_{d}e^{-i\omega_{d}t},
\nonumber\\
\dot{c}_{b}e^{-i\omega_{b}t}=i\Omega_{ab}e^{i\phi_{ab}}\cos\left(\nu t\right)c_{a}e^{-i\omega_{a}t}+i\Omega_{bd}e^{-i\phi_{bd}}\cos\left(\nu t\right)c_{d}e^{-i\omega_{d}t},
\nonumber\\
\dot{c}_{d}e^{-i\omega_{d}t}=i\Omega_{ad}e^{i\phi_{ad}}\cos\left(\nu t\right)c_{a}e^{-i\omega_{a}t}+i\Omega_{bd}e^{i\phi_{bd}}\cos\left(\nu t\right)c_{b}e^{-i\omega_{b}t}.\nonumber
\end{gather}
Since the pulse frequencies negligibly couple the two closely-spaced excited states we get
\begin{gather}
\dot{c}_{a}e^{-i\omega_{a}t}=i\Omega_{ab}e^{-i\phi_{ab}}\cos\left(\nu t\right)c_{b}e^{-i\omega_{b}t}+i\Omega_{ad}e^{-i\phi_{ad}}\cos\left(\nu t\right)c_{d}e^{-i\omega_{d}t},
\nonumber\\
\dot{c}_{b}e^{-i\omega_{b}t}=i\Omega_{ab}e^{i\phi_{ab}}\cos\left(\nu t\right)c_{a}e^{-i\omega_{a}t},
\nonumber\\
\dot{c}_{d}e^{-i\omega_{d}t}=i\Omega_{ad}e^{i\phi_{ad}}\cos\left(\nu t\right)c_{a}e^{-i\omega_{a}t}.\nonumber
\end{gather}
 We then multiply the last two equations by $\Omega_{ab},\Omega_{ad}$ and add them
to obtain
\begin{gather}
\Omega_{ab}e^{-i\phi_{ab}}\dot{c}_{b}+\Omega_{ad}e^{-i\phi_{ad}}\dot{c}_{d}=i(\Omega_{ab}^2e^{-i\left(\omega_{a}-\omega_{b}\right)t}+\Omega_{ad}^2e^{-i\left(\omega_{a}-\omega_{d}\right)t})\cos\left(\nu t\right)c_{a}
\nonumber\\
\dot{c}_{a}e^{-i\omega_{a}t}=i\Omega_{ab}e^{-i\phi_{ab}}\cos\left(\nu t\right)c_{b}e^{-i\omega_{b}t}+i\Omega_{ad}e^{-i\phi_{ad}}\cos\left(\nu t\right)c_{d}e^{-i\omega_{d}t}.\nonumber
\end{gather}

For short pulse duration compared to $1/|\omega_d-\omega_b|$ we have  $\left(v-\left(\omega_{a}-\omega_{b}\right)\right)t\ll1,\left(v-\left(\omega_{a}-\omega_{d}\right)\right)t\ll1,$ and
we write
\begin{gather}
\frac{\Omega_{ab}e^{-i\phi_{ab}}\dot{c}_{b}+\Omega_{ad}e^{-i\phi_{ad}}\dot{c}_{d}}{\Omega_T}=\Omega_T i\cos\left(\nu t\right)c_{a},\nonumber\\
\dot{c}_{a}=i\Omega_T\cos\left(\nu t\right)\frac{\left(\Omega_{ab}e^{-i\phi_{ab}}c_{b}+\Omega_{ad}e^{-i\phi_{ad}}c_{d}\right)}{\Omega_T}.\nonumber\\
\Omega_T=\sqrt{\Omega^2_{ab}+\Omega^2_{ad}}.\nonumber
\end{gather}
We can thus define an effective level 
\[
c_{e}\equiv \frac{\Omega_{ab}e^{-i\phi_{ab}}c_{b}+\Omega_{ad}e^{-i\phi_{ad}}c_{d}}{\Omega_T},
\]
and get a similar solution to the one obtained for the 2-level system with a temporally constant $\Omega_R$ \cite{scully1997quantum}. One can then express the polarization using this  effective level with an emission frequency equal to the difference between the frequencies of $\left|c_e\right>$ and $\left|a\right>$
$$P=\mu_1\left|a\right>\left<b\right|+\mu_2\left|a\right>\left<d\right|+h.c.=\mu_\mathrm{eff}\left|a\right>\left<c_e\right|+h.c.$$
Note that in the expression for the polarization \cite{scully1997quantum}, the phase of the dipole
matrix element cancels out and the emissions from $b$ and $d$ levels
have the same phase, which agrees with the effective two-level system description. While these emissions are at slightly different
frequencies, the beating pattern is negligible at short times. Similarly, the Rabi splitting during light-atom interaction is negligible since $t\ll 1/\Omega.$

\subsection{Atomic response from expanding and numerically solving the Bloch equations}

We consider Bloch equations in the rotating-wave approximation \cite{alekseev1992analytic}:
\begin{align}
\frac{dv}{dt} &= -\Gamma v(t) + \Omega f(t) n(t) \\
\frac{dn}{dt} &= -\Gamma n(t) - \Omega f(t) v(t)+n_0\Gamma
\end{align}
where $v(t)=2\mathrm{Im}(\rho_{12}),$ $f(t)$ is the incoming-field envelope and
the initial conditions are:
\[
v(0) = 0, \quad n(0) = n_0.
\]
For  \( f(t) =\exp(-\Gamma t)\) we expand to second order $n(t)=n_0+n_1t+\frac {1}{2}n_2t^2+\mathcal{O}(t^3),v(t)=v_1t+\frac {1}{2}v_2^2+\mathcal{O}(t^3),f(t)=1-\Gamma t+\frac{1}{2}\Gamma^2t^2+\mathcal{O}(t^3)$ and substitute them in the equations. By equating the same orders, we get exactly the same expression for $v(t)$ as in the main text $v(t)=\Omega n_0 t-\Gamma \Omega n_0 t^2+\mathcal{O}(t^3).$

To verify the integration operation we expand $v(t)$ for $f(t)=f_0+f_1t+...$  and obtain: 
$$v(t)=\Omega\left[f_0n_0t+\frac{1}{2}n_0(f_1-\Gamma f_0)t^2\right]+\mathcal{O}(t^3),$$
where $f_0\sim 1,f_1\gg \Gamma ,$ and we approximately get integration as expected.
Finally, we cross validate the integration operation by numerically solving the Bloch equations for $f(t)=\frac{1}{1+t^2}$ and comparing it  to the analytic result. As can be seen in Fig. 1 there is excellent agreement between $v(t)$ and the analytic integration result.

\begin{figure}[h]
    \centering
    \includegraphics[width=8cm]{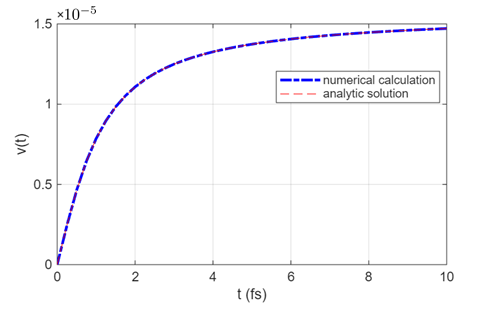}
    \caption{Comparison  of $v(t)$ for $f(t)=\frac{1}{1+at^2},$ where $a=10^{15} \,(\mathrm{Hz}),$ by numerically solving differential Bloch equations to the analytic integral $\frac{\Omega}{a} \mathrm{atan}(at)$ with very good agreement.  }
    \label{fig:MB}
\end{figure}

\subsection{Validation of the atomic response for $\Gamma=0$}
To further validate the result in Eq. (1), we utilize the textbook expression for the polarization from Ref. \cite{scully1997quantum} for $\phi=0,\,\Delta=0$ given by:
$$ P(t)=2\mathrm{Re}\left\{ i\frac{\Omega_{R}}{\Omega}\mathcal{P}_{ab}\cos(\Omega t/2)\sin (\Omega t/2)e^{i\nu t}\right\} $$
Taking the $t\rightarrow 0$ limit we get
$$ 
P(t)=\mathrm{Re}\left\{ i\Omega_{R}\mathcal{P}_{ab}  te^{i\nu t}\right\} =-\Omega_{R}\mathcal{P}_{ab}  t\sin(\nu t),$$ which agrees with our derivation in the main text.

\subsection{Validation with time dependent Schrödinger equation}

We solve time dependent Schrödinger equation \cite{cohen1986quantum} under the assumptions of a weak perturbation
i.e., $\omega_{ni}\gg\Omega,$ where $\omega_{ni}$ is the transition frequency, and $\Omega t \ll 1,$ enabling to use first-order perturbation theory. Employing the dipole approximation and the gauge transformation $\psi\left(\mathbf{r},t\right)=\exp\left[\frac{ie}{\hbar}\mathbf{A}\left(\mathbf{r}_{0},t\right)\cdot\mathbf{r}\right]\phi\left(\mathbf{r},t\right),$ the interaction Hamiltonian reads $H_{\rm int} = e \mathbf{r} \cdot \mathbf{E}(\mathbf{r}_0,t)$ \cite{scully1997quantum}. We consider a driving field of the form $\mathbf{E}(t)=\theta(t)\mathbf{E}_0 e^{i\omega t} f(t),$ where $f(t)$ is the field envelope, with circular polarization \cite{scully1997quantum}, which eliminates counter-rotating terms. We then calculate the final-state amplitude
\[
b_{n}^{\left(1\right)}\left(t\right)=\frac{W_{ni}}{i\hbar}\intop_{0}^{t} e^{i\left(\omega_{ni}- \omega\right)t'}f\left(t'\right)dt'=\frac{W_{ni}}{i\hbar}\intop_{0}^{t} f\left(t'\right)dt',
\]
where $W_{ni}=\left<\psi_f|e \mathbf{r} \cdot\mathbf{E}_0|\psi_i\right>$ is the matrix element for the transition between states without the temporal dependency of the field and we used a carrier frequency that equals the transition frequency $\omega=\omega_{ni}.$ For $t\ll1/\Omega$ the initial-state population $b_{i}\left(t\right)=\sqrt{1-\left|b_{n}^{\left(1\right)}\left(t\right)\right|^{2}}\sim 1-\frac{\left(W_{ni}t\right){}^{2}}{2\hbar^{2}}= 1-\frac{\left(\Omega t\right){}^{2}}{2} \approx  1.$ We thus get that $\rho_{21}$ is proportional to the pulse-envelope integral:
\[
\rho_{21}=b_{i}\left(t\right)\left(b_{n}^{\left(1\right)}\left(t\right)\right)^{*}\approx \frac{W_{ni}}{i\hbar}\intop_{0}^{t} f\left(t'\right)dt',
\]
in agreement with our Bloch-equation analysis. Note that this regime of approximately constant initial state population is the standard regime in transient absorption experiments \cite{goulielmakis2010real}.

\subsection{Relation between the temporal resolution and the dipole approximation}
For a general waveform with a carrier frequency $\omega_1$ and envelope $f(t)$, we get for the Fourier transform:
$$f\left(t\right)e^{i\omega_1 t}\rightarrow F\left(\omega\right)*\delta\left(\omega-\omega_{1}\right)=F\left(\omega-\omega_{1}\right).$$
Thus, the Fourier transform of the pulse envelope extends the frequency content relative to the carrier frequency. 
For concerteness, we consider a 10 fs Gaussian pulse with $\sigma_t=5$ fs and obtain
$$e^{-\alpha t^2}\rightarrow e^{-\frac{\omega^2}{4\alpha}},\sigma_t=5\,\mathrm{fs},\alpha=\frac{1}{(5\mathrm{fs})^2},\sigma_\omega=\frac{1}{2.5\mathrm{fs}} .$$
Hence, for $\lambda=$550 nm this translates  to angular frequency content up to $3.4\cdot 10^{15}\mathrm{(1/s)}+3\sigma_\omega=4.6\cdot 10^{15}\mathrm{(1/s)}.$
The dipole approximation permits wavelengths of 5 nm and longer corresponding to angular frequencies up to $\omega=3.77\cdot10^{17}$(1/s) and thus this is well within the dipole approximation. Processing such a Gaussian pulse requires approximately 50 time steps, which tranlsates to 200 attosecond resolution. Performing a similar calculation for a 500 attosecond Gaussian pulse with a carrier frequency of the He transition at 58.4 nm shows that the dipole approximation enables temporal resolution of 10 attosecond. 
In general, one can write
$\omega_c+\omega_\mathrm{e\, max}\ll \omega_{a_0},$
where $\omega_c$ is the angular carrier frequency, $\omega_\mathrm{e\, max}$ is the maximal envelope modulation frequency, $\omega_{a_0}=\frac{2\pi c}{a_0},$ and $a_0$ is the Bohr radius ($a_0$ approximates the wavefunction extent). This can be written as $\frac{ c}{\lambda}+\frac{1}{\mathrm{temporal\, resolution}}\ll \frac { c}{a_0}.$

\subsection{A scheme of the setup}
\begin{figure*}[!t]
    \centering
    \includegraphics[width=15cm]{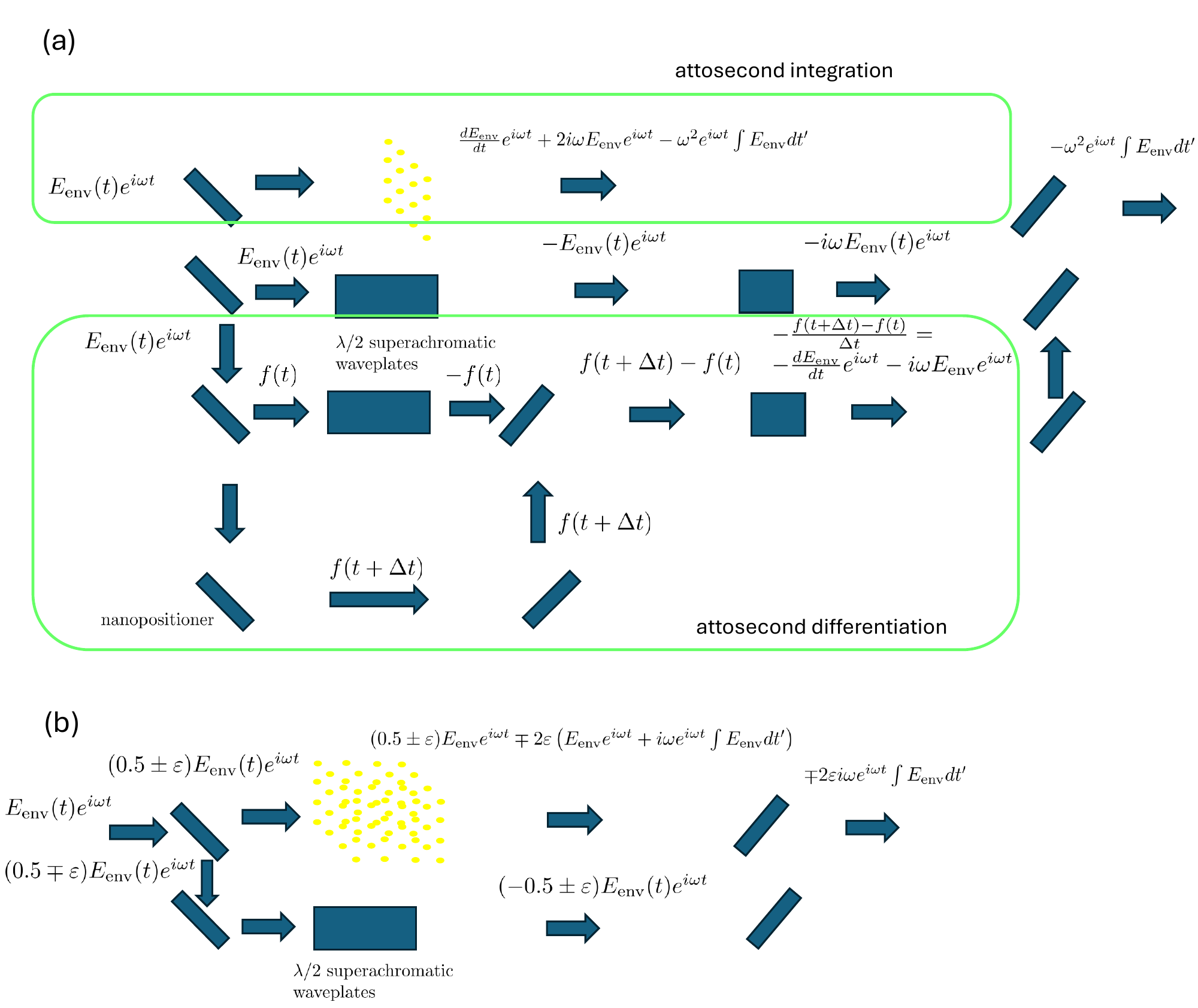}
    \caption{A scheme of the setup with a 2D atomic array and bulk realizations. (a) The incoming pulse impinges on a 2D atomic array, which results in adding or reducing a proportional-integration-differentiation signal via stimulated emission and absorption, respectively. The proportional and differentiation signals can be subtracted to obtain pure integration. The differentiation scheme can also be used independently. \hl{(b) The incoming pulse impinges on a bulk (e.g., gas cell), producing the integral and proportional signals. The proportional signal (incoming and generated) can be subtracted to obtain pure integration. Note that the differentiation part can be applied independently and any proportional-integral-derivative combination can be generated in this way.  } }
\label{fig:enter-labelscheme}
\end{figure*}
We present in Fig. 2 a scheme of the setup. The setup is composed of an atom array (a) or gas cell (b), which emits or absorbs a field. In Fig. 2 (a) this field is composed of proportional, integral, and derivative signals. We therefore split the incoming field in order to also generate the proportional and derivative signals. By interfering them, we can produce any combination in the output including pure integration. Note that this signal is generated on top of the incoming signal, which has to be subtracted as well. While regular atom arrangement with an atom separation on the order of the wavelength results in additional diffraction orders, which reduce the signal, when the atom spacing is irregular, it eliminates these additional diffraction orders, thereby increasing the signal. Finally, when the spacing is larger than the wavelength, the atoms behave approximately independently, enabling to maintain the atomic temporal resolution. \hl{In Fig. 2 (b) we present the gas cell setup. In such a setup the output signal interferes constructively and negligibly interacts with the atoms} \cite{goulielmakis2010real}. \hl{Specifically, the polarization interferes with the forward-propagating wave resulting in an integral and proportional signals}.  The Rb atom transitions D\textsubscript{1} and D\textsubscript{2} have a frequency difference of 7.12 THz, enabling the use of $200$ fs pulses.  Strontium atoms, which have no hyperfine structure, and possess well separated excited states allow for 10 fs pulses (recently trapped in an optical-tweezer arrays \cite{grun2024optical,norcia2018microscopic}).

\subsection{Practical atomic systems to realize the phenomenon}

In the D\textsubscript{2} transition of rubidium atoms, the ground and excited states have  splittings of 4.27 GHz and 0.3 GHz, respectively \cite{bize1999high}, enabling one to use pulses of up to 5 picosecond with the effective two-level description described above. Using optical pumping, the ground-state population of Rb atoms can be directed into a chosen hyperfine ground state, further increasing the maximal pulse length to 50 ps. The frequency difference between the D\textsubscript{1} and D\textsubscript{2} excited states is 7.12 THz, setting a minimum pulse width to 200 fs to selectively excite one level. Interestingly, $^{166}\mathrm{Er}$ and Sr atoms have nuclear spin $I=0$ \cite{grun2024optical} and therefore no hyperfine structure, eliminating this splitting type. In particular, Sr has the single dominant transition $5s^2\, {}^1S_0 \longrightarrow  5s5p\, {}^1P_1$ at 461 nm, with the closest weaker transition from the ground state $5s^2\; {}^1S_0 \to 5s5p\; {}^3P_1$ at 689 nm, enabling also high-modulation frequencies e.g., 10 fs pulses will target mainly this transition \cite{Bertrand2021SrCooling,NIST_Sr_Table3,morton2000atomic,fuhr1996nist} (the electronic wavefunction extent is 2~\AA~corresponding to approximately 10 as resolution, which in turn allows for using pulses of 1 fs or longer). As this secondary transition is weaker by $\sim$4000, even this 10 fs restriction is not required, positioning Sr as an excellent atom for this approach. Helium also has no nuclear spin and well-separated transitions, with the dominant $1s^2\,{}^1S_0\longrightarrow 1s2p\,{}^1P_1$ at 58.4 nm, allowing for sub femtosecond XUV pulses  \cite{martin1960energy,Martin2002_HeI,kneller2025attosecond}. Specifically, the closest transition is the $1s^2\,{}^1S_0\longrightarrow 1s3p\,{}^1P_1$ at 53.7 nm, enabling to use pulses longer than 800 as for selective excitation (the electron wavefunction extent  is 0.8~\AA~corresponding to 5 attosecond resolution, which in turn allows  for using such pulses). Note that circular polarization can drive the above transitions. Finally, a larger spectral separation between transitions—sufficient to support pulse durations of 200 as—can be achieved using hydrogen-like ions such as $\mathrm{He^+}$ (30.37 nm, 25.63 nm) \cite{NIST_ASD}.
\subsection{Temporal response of a shorter slab}
\begin{figure}[t]
    \centering
    \includegraphics[width=8cm]{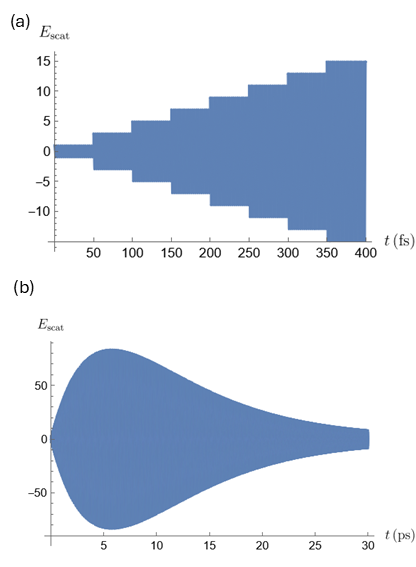}
    \caption{The response of a lossless slab to a complex-frequency resonant excitation. (a) The initial response of a slab with $l=10\lambda, r_1=0.99,\,\, n_1=1.4,\lambda=550\mathrm{nm},\,\,\mathrm{and}\,\, Q=2810$ to the complex-resonant-frequency  incoming field $E_\mathrm{inc}=\exp(-i\omega t -\Gamma t)\theta(t).$ Unlike the previous case, here the long roundtrip results in  discretization in the response, which significantly limits the temporal resolution. (b) The scattered field for the complex-frequency excitation for  $t<4/\Gamma,$  of the form $E_\mathrm{scat}\propto t\exp(-i\omega t -\Gamma t)\theta(t).$}
    \label{fig:enter-label3}
\end{figure}

Here we consider a short slab with the goal of improving the temporal resolution at the cost of much shorter operation time. To that end we choose the following parameters: $l_1=10\lambda,\,\,r_1=0.99,\,\,n=1.4,\,Q\,\,\mathrm{factor}=2810,\,\,\mathrm{and}\,\lambda=550\mathrm{nm}.$ In Fig. 3 (a) we plot the scattered field for an incoming field with a complex-frequency excitation of the form $e^{-\Gamma t-i\omega t}$ for   $t\ll 1/\Gamma$. This time the response is composed of steps with a time difference of the roundtrip of the slab of $\Delta t=51.3\mathrm{fs,}$ which is relatively long even for this  short slab. Similarly, when we input the same pulse as in Fig. 2 (d) (main text) of $e^{i\omega t}\frac{1}{1+t^2}$ for 10fs, we get that the scattered field is the incoming field multiplied by $r_1$ i.e., $r_1e^{i\omega t}\frac{1}{1+t^2}$ (did not perform any integration) instead of its integral during this pulse. In Fig. 3 (b) we plot the response of the slab for longer times. Clearly, at longer times $E_\mathrm{scat}$ qualitatively behaves  similarly to the response of the classical subwavelength passive resonators considered in Ref. \cite{farhi2025time} and an atom with $\Gamma>\Omega$, with a $te^{-\Gamma t}$ scaling. Here the linear behavior lasts for more optical cycles compared to the subwavelength particle due to the higher Q factor of the slab but for significantly less optical cycles than the atom.

\end{widetext}
\end{document}